\documentclass[aps,pre,floats,superscriptaddress,showpacs,twocolumn]{revtex4}

\usepackage{amsmath, amsthm, amssymb}
\usepackage{enumitem}
\usepackage{graphics,graphicx}% Include figure files
\usepackage{epsfig}
\usepackage{times}
\usepackage{dcolumn}% Align table columns on decimal point
\usepackage{bm}% bold math

\usepackage{ulem, color}

\begin{document}

\title{Generalized synchronization in relay systems with instantaneous coupling}

\author{R. Guti\'{e}rrez}
\affiliation{Department of Chemical Physics, The Weizmann Institute of Science, Rehovot 76100, Israel}
\author{R. Sevilla-Escoboza}
\affiliation{Centro Universitario de los Lagos, Universidad de Guadalajara, Lagos de Moreno, Jalisco 47460, M\'exico}
\affiliation{Complex Systems Group, Universidad Rey Juan Carlos, 28933 M\'ostoles, Madrid, Spain}
\author{P. Piedrahita}
\affiliation{Institute for Biocomputation and Physics of Complex Systems (BIFI), University of Zaragoza, Zaragoza 50009, Spain}
\author{C. Finke}
\affiliation{d-fine GmbH, Opernplatz 2, 60313 Frankfurt, Germany}
\author{U. Feudel}
\affiliation{ICBM, University of Oldenburg, Carl-von-Ossietzky-Strasse 9-11, 26111 Oldenburg, Germany}
\affiliation{IPST, University of Maryland, College Park, MD 20742-2431, USA}
\author{J. M. Buld\'u}
\affiliation{Complex Systems Group, Universidad Rey Juan Carlos, 28933 M\'ostoles, Madrid, Spain}
\affiliation{Center for Biomedical Technology, Technical University of Madrid, Pozuelo de Alarc\'{o}n, 28223 Madrid, Spain}
\author{G. Huerta-Cuellar}
\affiliation{Centro Universitario de los Lagos, Universidad de Guadalajara, Lagos de Moreno, Jalisco 47460, M\'exico}
\author{R. Jaimes-Re\'ategui}
\affiliation{Centro Universitario de los Lagos, Universidad de Guadalajara, Lagos de Moreno, Jalisco 47460, M\'exico}
\author{Y. Moreno}
\affiliation{Institute for Biocomputation and Physics of Complex Systems (BIFI), University of Zaragoza, Zaragoza 50009, Spain}
\affiliation{Department of Theoretical Physics, University of Zaragoza, Zaragoza 50009, Spain}
\affiliation{Complex Networks and Systems Lagrange Lab, Institute for Scientific Interchange, Turin, Italy}
\author{S. Boccaletti}
\affiliation{CNR- Institute of Complex Systems, Via Madonna del Piano, 10, 50019 Sesto Fiorentino, Florence, Italy}

\date{\today}

\begin{abstract}
We demonstrate the existence of generalized synchronization in systems that act
as mediators between two dynamical units that, in turn, show complete synchronization with each other. These are the so-called relay systems. Specifically, we analyze the Lyapunov spectrum of the full system to elucidate when complete and generalized synchronization appear. We show that once a critical coupling strength is achieved, complete synchronization emerges between the
systems to be synchronized, and at the same point, generalized synchronization with the relay system also arises. Next, we use two nonlinear measures based on the distance between phase-space neighbors to quantify the generalized synchronization in discretized time series. Finally, we experimentally show the robustness of the phenomenon and of the theoretical tools here proposed to characterize it.
\end{abstract}


\maketitle

Synchronization is a common phenomenon in a diversity of natural and technological systems \cite{pikovsky2001}. Synchrony, however, is not always achieved spontaneously, and reaching or maintaining a synchronous state often requires an external action. 
%In particular, a successful way of synchronizing oscillators with moderate to low couplings is through external driving with both deterministic and stochastic signals.
An elegant way to enhance synchronization is the use of relay units between the systems to be synchronized (see Fig. 1a). Relay synchronization (RS) consists in achieving complete synchronization (CS) of two dynamical systems by indirect coupling through a relay unit, whose dynamics does not necessary join the synchronous state. RS is especially useful in bidirectionally coupled systems with a certain delay in the coupling line. In these cases, indeed, the coupling delay may induce instability of the synchronous state \cite{heil2001}, which can be restored again thanks to a relay system.
Lasers \cite{fischer2006} and electronics circuits \cite{wagemakers2007}
have been the benchmark for experimental demonstration of the feasibility of RS, showing its robustness against noise or parameter mismatch. In semiconductor lasers, for instance, zero-lag synchronization between two delay-coupled oscillators can be achieved by relaying the dynamics via a third mediating element, which surprisingly lags behind the synchronized outer elements. With electronic circuits, RS has been used as a technique for transmitting and recovering encrypted messages, which can be sent bidirectionally and simultaneously \cite{wagemakers2008}. Apart from its technological applications, RS has also been proposed as a possible mechanism at the basis of isochronous synchronization between distant areas of the brain \cite{engel1991}. Despite such evidence of RS, there are still open questions of a fundamental nature. The main issue is to characterize properly the relationship, established in RS, between the dynamics of the relay system and that of the synchronized systems. When a certain delay is introduced in the coupling lines, lag-synchronization has been reported \cite{fischer2006}. Nevertheless, relay
units may have certain parameter mismatch \cite{banerjee2012} or even be completely different systems \cite{wagemakers2007}, thus having dynamics with unclear a priori relationship with the systems they are synchronizing.

In this paper, we give evidence that RS in fact corresponds to the setting of generalized synchronization (GS) between the relay system and the synchronized systems. Given two dynamical systems whose dynamics are given, respectively, by $\dot x(t)=f(x(t),y(t))$ and $\dot y(t)=g(y(t),x(t))$, GS is based on the existence of a one-to-one function $h(x(t))$ such that $\lim_{t \to \infty} \lVert y(t)- h(x(t)) \rVert =0$ \cite{pikovsky2001}.
The existence of GS in {\it unidirectionally} coupled units (drive system $\rightarrow$ response system) has been proven by checking the ability of the response system to react identically to different initial conditions of the same driver system, which can be quantified by
evaluating the mutual false nearest neighbors \cite{rulkov1995} or by measuring the conditional Lyapunov exponents \cite{kocarev1996}. Recently, GS has been also reported in networks of {\it bidirectionally} coupled oscillators \cite{moshkalenko2012}. While recently it has been suggested that GS could occur when a {\it minimum} value of the coupling delay is guaranteed in a relay configuration \cite{landsman}, no proof of GS existed so far for systems that are {\it instantaneously} coupled through an additional relay unit.

\begin{figure}[!bht]
\includegraphics[scale=0.40]{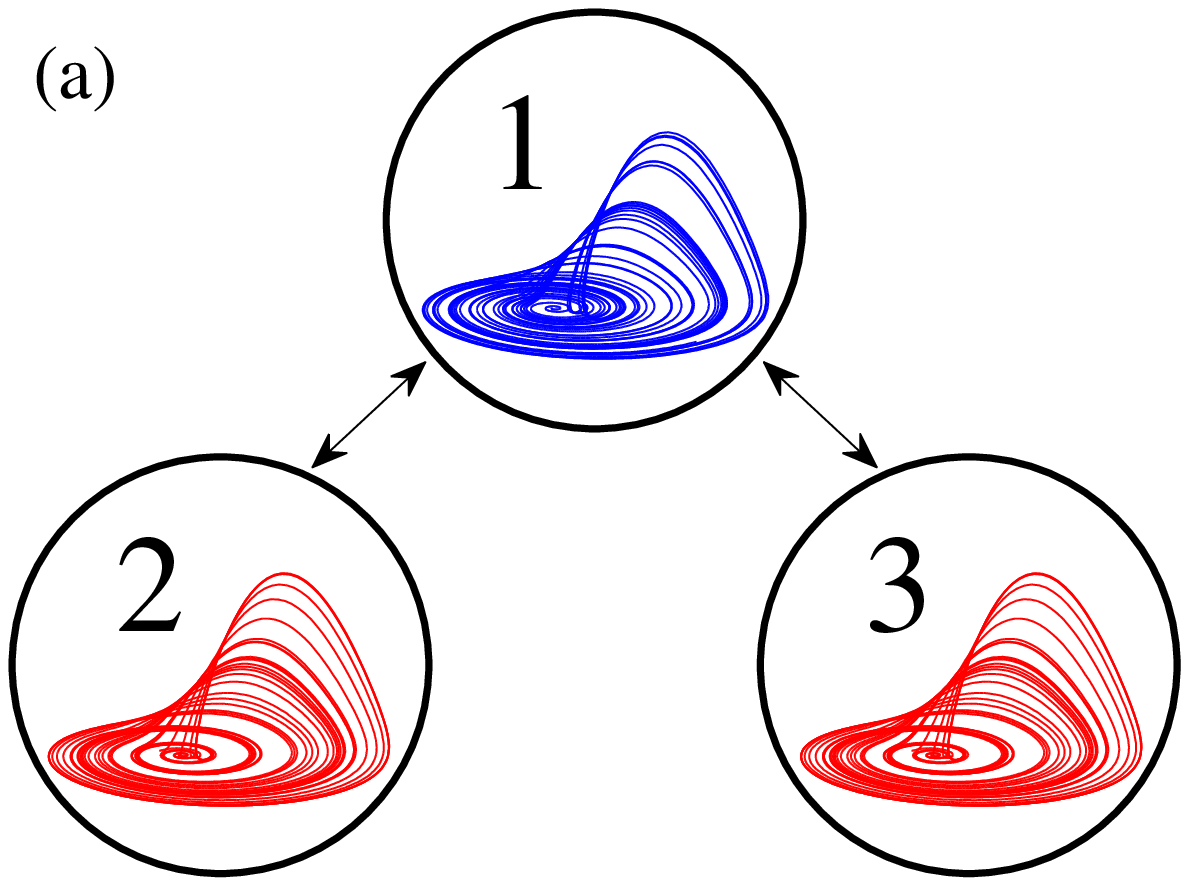}
\includegraphics[scale=0.53]{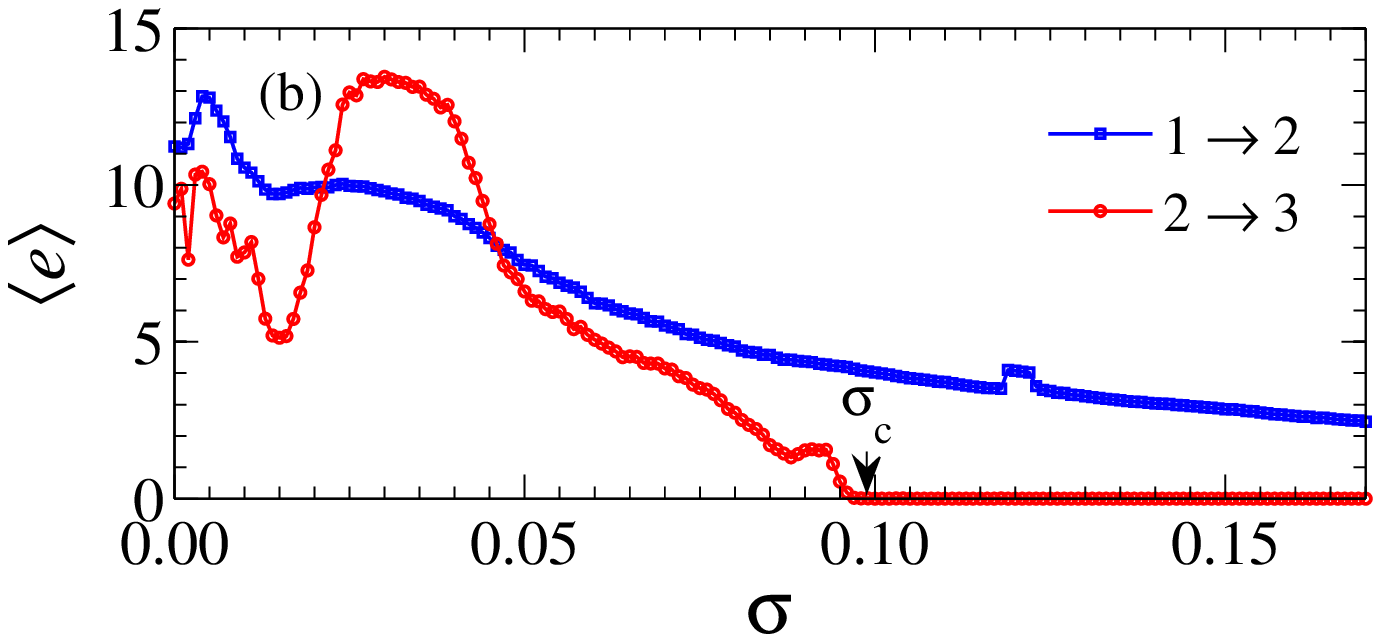}
\caption{\label{fig1} (Color online). (a) Relay configuration scheme of the system of Eqs. (\ref{rossler}). (b) Synchronization error $\langle e\rangle$ (see text for definition) between systems 1 and 2 (blue line), and between systems 2 and 3 (red line) as a function of the coupling strength $\sigma$. The critical coupling $\sigma_c$ marks the beginning of the RS regime.}
\end{figure}

With the aim of determining whether GS is behind the role played by the relay system, we start by considering the case of three interacting R\"ossler oscillators \cite{rossler} diffusively coupled according to the configuration scheme of Fig.~\ref{fig1} (a). The generic route to complete synchronization of two R\"ossler oscillators is well known in the literature \cite{pikovsky2001}. Here, instead, we consider a relay configuration in which oscillators 2 and 3 are identical, whereas oscillator 1 (the relay unit) is set to have (one or more) different parameters with respect to them. The coupling is assumed to be bidirectional and instantaneous. RS is said to occur whenever complete synchronization (CS) between oscillators 2 and 3 is observed.

The equations of motion of the full system are
\begin{equation}\label{rossler}
\begin{aligned}
&\left\{
\begin{aligned}
&\dot{x}_1 = - y_1 - z_1,\\
&\dot{y}_1 = x_1 + a_0\, y_1 + \sigma (y_2 - y_1) + \sigma (y_3 - y_1),\\
&\dot{z}_1 = 0.2 + z_1(x_1-5.7),\\
\end{aligned}
\right.\\
&\left\{
\begin{aligned}
&\dot{x}_{2,3} = - y_{2,3} - z_{2,3},\\
&\dot{y}_{2,3} = x_{2,3} + a\, y_{2,3} + \sigma (y_1 - y_{2,3}),\\
&\dot{z}_{2,3} = 0.2 + z_{2,3}(x_{2,3}-5.7).\\
\end{aligned}
\right.
\end{aligned}
\end{equation}

We focus on the case  $a_0 = 0.3$, $a = 0.2$, although different parameter mismatches between unit 1 and units 2 and 3 have also been tested with the same qualitative results. In all cases considered, the existence of a stable chaotic attractor has been verified for the isolated systems \cite{integration}. Additionally, the synchronization error $\langle e\rangle_{i,j}$ between the units $i$ and $j$ is defined as $\lim_{\tau\to \infty} \tau^{-1} \int_0^\tau \lVert {\bf x}_i(t) - {\bf x}_j(t) \rVert dt$. Figure~\ref{fig1} (b) shows  $\langle e\rangle_{1,2}$ (blue line) and $\langle e\rangle_{2,3}$  (red line) \cite{syncerror}, as a function of the coupling strength $\sigma$. It is clear that there is a critical value for the coupling, $\sigma_c \simeq 0.10$, above which RS occurs for any generic initial condition, where complete synchronization between units 2 and 3 occurs, whereas the relay system 1 still displays $e_{1,2}>0$.

\begin{figure}[!bht]
\includegraphics[width=\columnwidth]{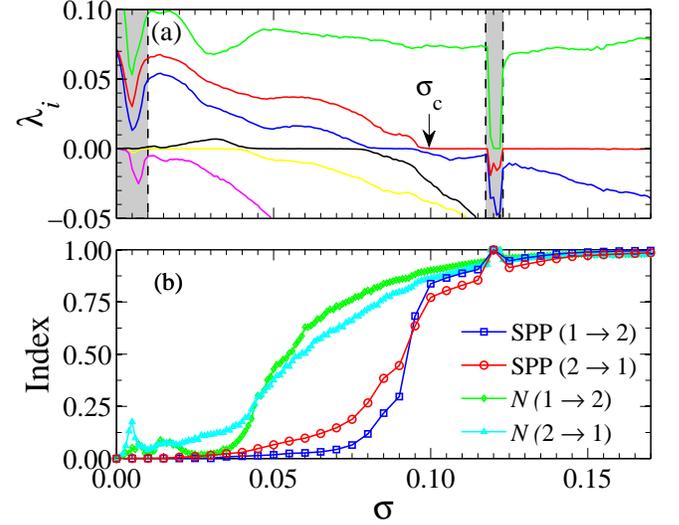}
\caption{\label{fig2} (Color online). (a) Lyapunov spectrum  as a function of  $\sigma$ (only the positive, zero and slightly negative Lyapunov exponents are shown). The gray filled areas are windows where the dynamics is periodic or only slightly chaotic (see the text for further explanations). The critical coupling $\sigma_c$ coincides to a very high precision with the coupling strength at which the second largest Lyapunov exponent vanishes.  (b) SPP and $N$-index  (see text for definition) vs. $\sigma$, with system 1 as the domain set and 2 as the codomain set for the possible mapping (blue line for SPP and green line for $N$-index) and vice-versa (red line for SPP and light blue line for $N$-index).}
\end{figure}

More insight into the role that the relay system plays in RS is gained by computation of the Lyapunov spectrum of the full 9-dimensional system, which is here realized by means of the classical method by Benettin et al. \cite{benettin,lyapunov}. The results are reported in Fig.~\ref{fig2} (a), where the 6 largest Lyapunov exponents in the spectrum are plotted as a function of $\sigma$. The highlighted areas are windows where periodic dynamics show up in certain realizations (in the case of $\sigma \simeq 0.006$), or in all of them (around $\sigma \simeq 0.12$).  Consequently, we do not consider any further these coupling regions where simpler dynamical regimes accidentally emerge, as they do not add relevant information for the understanding of RS.

%%%%%%%%%
%%%%%%%%%%%%

For negligible couplings, the set of Lyapunov exponents (labeled such that $\lambda_1 \geq \lambda_2 \geq \cdots \geq \lambda_9$)  is divided into three positive ($\lambda_1 > 0$ and $\lambda_2 = \lambda_3 >0$), three zero ($\lambda_4 = \lambda_5 = \lambda_6 = 0$), and three negative ($\lambda_7 < 0$ and $\lambda_8 = \lambda_9 < 0$) exponents. As the coupling increases, $\lambda_6$ becomes negative almost immediately. By checking the phase-space orbits of the systems, this corresponds to a phase synchronization regime between systems 2 and 3. At $\sigma \simeq 0.04$, $\lambda_5$ also becomes negative, and just one effective phase remains in the system, corresponding to $\lambda_4 = 0$. So far the three largest Lyapunov exponents remain positive, suggesting that the three chaotic amplitudes are still not correlated. A further increase in $\sigma$ determines the vanishing of $\lambda_3$ and the dropping below zero of $\lambda_4$.
Eventually, for higher coupling strengths, $\lambda_2$ vanishes and $\lambda_3$ becomes negative. The coupling strength for which this latter scenario is observed is $\sigma = 0.100 \pm 0.001$, and therefore it almost perfectly matches the critical coupling strength for RS. In other words, the onset of RS corresponds to a regime with only one independent chaotic amplitude in the entire system. The fact that $\lambda_1 > 0$, $\lambda_2 = 0$ and $\lambda_i < 0$ for $i=3,4,\ldots,9$ hints at the possibility that GS is taking place between any of the systems 2 or 3 (which are in complete synchronization) and system 1 (i.e., the possibility that there is a functional relationship  ${\bf x}_{2,3}(t) = h({\bf x}_1(t))$, and the phase-space trajectories collapse onto a generalized synchronization manifold).

Direct evidence of the onset of GS between systems 1 and 2 can be provided by the use of two indices (among others): the synchronization points percentage (SPP, introduced by Pastur {\it et al.} in Ref.~\cite{pastur}), and the N-index \cite{quianquiroga}. Briefly, SPP quantifies the fraction of phase-space points of a given subsystem for which there is a {\it local continuous function} to the phase-space of the other subsystem. The essence of the method is analyzing the nearest neighbors of the points in the domain subsystem, and looking at their images in the neighborhoods of time-related points in the codomain subsystem, this way asserting the existence of  local functions only for certain statistical confidence level (continuity statistics method) \cite{pecora}. A way to optimize this search is  performing the so-called time-delay reconstruction of the subspaces involved \cite{kennel}, due to the fact that, in higher dimensions, the size of these neighborhoods (the number of points inside co-domains required to assess the existence of the local function) is smaller. Note that, even though this reconstruction is convenient in terms of time efficiency, it is just an optional step before SPP computation. Whenever SPP = 1, there exists a unique, global, continuous synchronization function from one subsystem to the other \cite{bocca}, and thus we say that the two subsystems are in GS (see Ref.~\cite{pastur} for further details of the method). The second index used is the N-index, a nonlinear measure of synchronization proposed in Ref.~\cite{quianquiroga}, which is defined as

\begin{equation}
N({\bf x}|{\bf y}) = \frac{1}{P}\sum_{n=1}^P \frac{R_n({\bf x}) - R_n^{(k)}({\bf x}|{\bf y})}{R_n({\bf x})},
\end{equation}

\noindent where ${\bf x}(t)$ and ${\bf y}(t)$ are the states of the two dynamical systems for which GS is being evaluated, and the subindex $n=1,...,P$ refers to a discrete-time sampling of the attractor. Furthermore, $R_n({\bf x}) = (P-1)^{-1} \sum_{i\neq n} ({\bf x}_n - {\bf x}_i)^2$ is the mean squared distance to random points in the attractor, and $ R_n^{(k)}({\bf x}|{\bf y}) = k^{-1} \sum_{i=1}^k ({\bf x}_n - {\bf x}_{{\bf y}_{n,i}})^2$ is the mean squared distance to the $k$ false nearest neighbors of ${\bf x}_n$, which are the points corresponding to the time indices ${\bf y}_{n,i}$ of the $k$ nearest neighbors of ${\bf y}_n$. By definition, $N({\bf x}|{\bf y}) \leq 1$, and it can be marginally smaller than 0 for totally unsynchronized dynamics. Values close to zero indicate that there is no synchronization, whereas values close to 1 reflect the fact that for any $n$ a small cloud of neighboring points around ${\bf y}_n$ is mapped into a small cloud of neighboring points around ${\bf x}_n$, which hints again at
the presence of GS in the system (as it indicates the existence of a continuous mapping from the phase space of system ${\bf y}(t)$ to that of system ${\bf x}(t)$).

Figure~\ref{fig2} (b) shows the curves of both the SPP \cite{SPP} and the $N$-index  \cite{N}, for the case in which system 1 (2) is taken as reference and system 2 (1) is inspected for the existence of a functional relationship (denoted by  $1\rightarrow 2$ ($2\rightarrow 1$) in the Figure). The SPP curves clearly display a smooth behavior for almost every $\sigma$, and exhibit the transition to GS near $\sigma_{c}$, detecting the periodic dynamics at $\sigma=0.12$ (discontinuous jump to SPP $= 1.0$). The curves of the $N$-index fluctuate slightly above zero for small couplings, while they reveal a clear monotonous growth with $\sigma$ beyond $\sigma \simeq 0.04$. At $\sigma = \sigma_c$, the $N$-index values are very close to $0.90$, and for higher coupling strengths they increase up to $0.98$ for $\sigma = 0.17$, the changes being, from this point on,  almost indistinguishable from numerical fluctuations. All this evidence confirm that a GS regime is associated to the setting of RS, with the function relating the states of the peripheral and relay units being invertible, which is not the general case of GS in
unidirectionally coupled systems \cite{rulkov1995}.

\begin{figure}[!bht]
\centering
\includegraphics[width=0.4\textwidth]{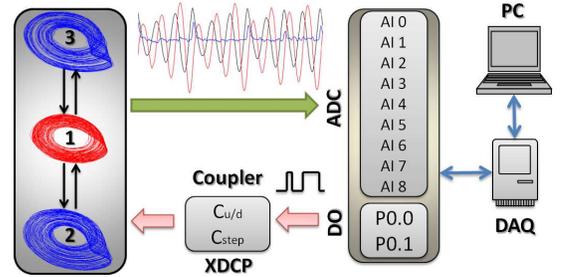}
\caption{\label{fig3} (Color online). Schematic representation of the experimental setup. The bidirectional coupling is adjusted by means of three digital potentiometers X9C104 (Coupler-XDCP) whose parameters $C_{u/d}$ (Up/Down Resistance) and $C_{step}$ (increment of the resistance at each step) are controlled by a digital signal coming from a DAQ Card. See text for the full details of the experimental system.}
\label{Lhop}
\end{figure}

Finally, we offer an evaluation of the robustness of these phenomena under realistic conditions, and we implement an experiment based on oscillating electronic circuits. The experimental setup is sketched in Fig.~\ref{fig3} and consists of three piecewise R\"{o}ssler circuits operating in a chaotic regime. The equations of motion of the experimental system are \cite{CP}:
\begin{equation}
\begin{aligned}
&\left\{
\begin{aligned}
&\dot{x}_1 = -\alpha \left( \Gamma\, x_1+\beta\, y_1+\xi\, z_1 -\sigma(x_{3} - x_{1}) -\sigma(x_{2}-x_{1})\right),\\
&\dot{y}_1 = -\alpha \left( -x_1+\upsilon_a\, y_1\right),\\
&\dot{z}_1 = -\alpha \left( -g\left( x_1\right) +z_1\right),\\
\end{aligned}
\right.\\
&\left\{
\begin{aligned}
&\dot{x}_{2,3} = -\alpha \left( \Gamma\, x_{2,3}+\beta\, y_{2,3}+\xi\, z_{2,3} -\sigma(x_1-x_{2,3}) \right),\\
&\dot{y}_{2,3} = -\alpha \left( -x_{2,3}+\upsilon_b\, y_{2,3}\right),\\
&\dot{z}_{2,3} = -\alpha \left( -g\left( x_{2,3}\right) +z_{2,3}\right).\\
\end{aligned}
\right.
\end{aligned}
\end{equation}
where the piecewise part is:
\begin{equation}
g(x_{i})= \left\{
\begin{aligned}
&0 & \textrm{if}\ & x_{i}\leq 3 \\
&\mu \left( x_{i}-3\right) & \textrm{if}\ & x_{i}>3%
\end{aligned}\right.
\end{equation}

\noindent Here, $\alpha=10^4\, \textrm{s}^{-1}$ is a time factor, and the other parameters are: $\Gamma=0.05$, $\beta=0.5$, $\xi=1$, $\mu=15$ and  $\upsilon_{a,b} =\frac{10}{R_{a,b}}-0.02$. The resistance mismatch ($R_a= 70\,\textrm{k}\Omega$, $R_b = 39\,\textrm{k}\Omega$) accounts for the difference between system 1 and systems 2 and 3, the latter being identical (this time, however, only up to tolerances of the electronic components and noise). The coupling strength $\sigma$ is controlled by a digital potentiometer (used as a voltage divisor), whose range is such that $\sigma \in \{0.00,0.01,\ldots,0.25\}$. We use three digital potentiometers (X9C104) which guarantee that the parameter $\sigma$ is changed simultaneously for all nodes. They are adjusted by a digital signal coming from ports P0.0 and P0.1 of a NI Instruments DAQ Card (DAQ). The output of each circuit is connected to a voltage follower that works as a buffer. All 9 signals are acquired by the analog ports (AI 0 ; AI 1; ... ; AI 8) of the same DAQ Card,
and recorded on a PC for further analysis.
The incoming signal of the analog inputs (ADC) and the signal sent through the digital outputs
(DO) are controlled and recorded by Labview Software.

\begin{figure}[!bht]
\includegraphics[width=\columnwidth]{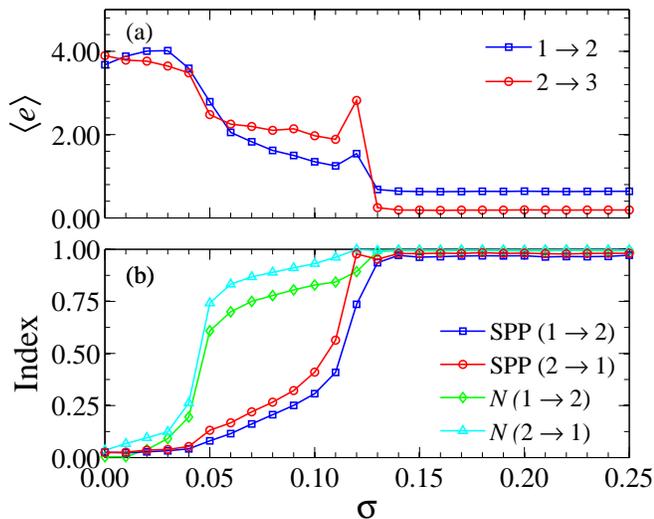}
\caption{\label{fig4} (Color online). (a) Synchronization error between systems 1 and 2 (blue line) and between systems 2 and 3 (red line) as a function of $\sigma$. (b) SPP and $N$-index (see text for definitions) vs. $\sigma$, with the same color stipulations as in the caption of Fig. 2 (b).}
\end{figure}

Figure~\ref{fig4} shows the values for the synchronization error (top), and the SPP and $N$-index (bottom) as functions of $\sigma$ for the experimental data. In particular, panel (a) indicates that the system achieves RS for $\sigma > 0.13$. Admittedly, the synchronization error $\langle e\rangle$ between systems 2 and 3 can never vanish, not even within experimental error limits in a low-precision experimental setup. However, it becomes very low as compared to  the considerably higher values  of $\langle e\rangle_{1,2}$. On the other hand, Fig.~\ref{fig4}(b) confirms that both SPP and $N$-index give clear indication on the existence of GS between systems 2 and 3 and system 1 for this experimental setup. The critical coupling observed in the synchronization error curves again matches very well with the point
where SPP and $N$-index become very close to 1, confirming the appearance of GS.

In summary, we have studied and characterized, both numerically and experimentally, the transition to synchronization of two chaotic systems when a third mediating unit acts, instantaneously, as the relay between them. We have demonstrated that relay synchronization can be associated to generalized
synchronization between the relay unit and the synchronized systems. The mediating role of
GS implies the existence of an invertible function
that links the dynamics of the relay system with those of the systems
to be synchronized. The key role of GS is demonstrated by analyzing the
Lyapunov spectrum of the whole system, the SPP and the $N$-index. Furthermore, the implemented electronic version of the coupled system shows the robustness of the results despite the inherent presence of noise and parameter mismatch.
Therefore, our results link the emergence of relay synchronization in instantaneously coupled chaotic systems with the existence of generalized synchronization with the relay system,
and open the possibility of using relay units for secure communications
\cite{uchida}. As recently demonstrated, indeed, chaos encryption by means of relay systems can be successfully implemented in real systems \cite{wu} and understanding the role of the relay unit will be fundamental for the feasibility of this kind of secure communications.

Authors acknowledge the computational resources and assistance provided by CRESCO, the center of ENEA in Portici, Italy. R.S.E. acknowledges Universidad de Guadalajara, Culagos (Mexico) for financial support through project PROINPEP 2012, Acuerdo No. RGS/013/2012, Subprograma 1, and Becas Mixtas 2012-2013  No. 290674 CVU 386032. C.F. and U.F. acknowledge financial support from Deutsche Forschungsgemeinschaft (FE 259/9) and the Burgers Program for Fluid Dynamics of the University of Maryland (U.F.).
Financial support  from MINECO (Spain) under projects FIS2011-25167, FIS2009-07072, of Comunidad de Madrid (Spain) under project MODELICO-CM S2009ESP-1691, and of the European Commission through the FET project MULTIPLEX (Grant 317532) is also acknowledged.


\begin{thebibliography}{}
\bibitem{integration}
For all numerical integration, a 4th-order Runge-Kutta algorithm with an integration time step of 0.005 time units has been used.
\bibitem{syncerror}
The synchronization error is computed for $2 \times 10^4$ time units, after an initial transient of $2 \times 10^3$ time units is disregarded. The curves in Fig. 1(b) refer to ensemble averages over 30 independent realizations, each one starting from a different random initial condition.
\bibitem{lyapunov}
For the computation of the Lyapunov exponents, a time window of $2 \times 10^4$ time units is considered after an initial transient of $2 \times 10^3$ time units. Gram-Schmidt reorthonormalization is applied every 0.25 time units. The reported curve are averages over 30 independent realizations, each one starting from a different random initial condition.
\bibitem{SPP}
A time window of $5 \times 10^5$ time units is considered after an initial transient of $5 \times 10^{4}$ time units. By keeping one 3D-point every $100$ integration steps, the resolution of the time series is $dt=0.5$ and the total number of points in each atttactor is $10^{6}$. For the time-delayed vectors (reconstruction) we choose embedding dimension $d=7$ and time delay $T=1$. The parameters of SPP algorithm are the number of points inside codomain neighborhoods $n=300$ (fixed over the attractor) and the threshold for statistical comparison $\Theta<0.1$.
\bibitem{N}
A time window of $10^5$ time units is considered after an initial transient of $2 \times 10^3$ time units. Even though the integration is performed with a time resolution of 0.005, only one time point every 0.5 time units is considered in the computation, so as to avoid temporal correlations. Finally, the computation of $N$-index is performed over the resulting $2 \times 10^5$ phase-space points. The procedure is repeated 30 times from random initial conditions.
\end{thebibliography}
\end{document}